\documentclass[aps,prl]{revtex4}
\def\ee{\end{equation}}
\def\bea{\begin{eqnarray}}

\def\bra#1{\langle #1 |}
\def\ket#1{| #1\rangle}

\def\Tr{{\rm Tr}}

\def\Prob{{\rm Prob}}
\usepackage{graphicx}
\DeclareGraphicsExtensions{.png}
\begin{document}

\title{Testing Causal Quantum Theory} 
\author{ Adrian Kent}
\email{A.P.A.Kent@damtp.cam.ac.uk} 
\affiliation{Centre for Quantum Information and Foundations, DAMTP, Centre for
Mathematical Sciences,
University of Cambridge, Wilberforce Road, Cambridge, CB3 0WA, United Kingdom}
\affiliation{Perimeter Institute for Theoretical Physics, 31 Caroline Street
North, Waterloo, ON N2L 2Y5, Canada.}

\begin{abstract}
Causal quantum theory assumes that measurements or collapses 
are well-defined physical processes, localised in space-time,
and never give perfectly reliable outcomes and that 
the outcome of one measurement only influences the outcomes
of others within its future light cone. 
Although the theory has unusual properties, it is 
not immediately evident that it is inconsistent with 
experiment to date.   I discuss its implications
and experimental tests. 
\end{abstract}
\maketitle
\section{Introduction}

There are fairly compelling theoretical and experimental reasons \cite{bell1966js,bell1989einstein,
bell1976theory,Pearle70,RKMSIMW01,MMMOM08,GMRWKBLCGNUZ13,ADR82,WJSWZ98,TBZG98,GZ99,
barrett2002quantum,gill2003accardi,hensen2015loophole,shalm2015strong,giustina2015significant,
SBHGZ08} to believe that nature violates local causality and that
local hidden variable theories are incorrect. 
However, while loopholes remain, the case is not quite conclusive.
In particular, the collapse locality loophole \cite{kent2005causal} remains
largely untested.  Moreover, causal quantum theory
\cite{kent2005causal} is a theoretically interesting
example of a local hidden variable theory that
remains consistent with Bell experiments to date by
exploiting this loophole.  

Causal quantum theory assumes that measurements or collapses 
are well-defined physical processes, localised in space-time,
and never give perfectly reliable outcomes.
Unlike standard quantum theory, it also assumes that measurement outcomes respect a strong form
of Minkowski or Einstein causality, in that the outcome of
one measurement only affects the probability distributions for
the outcomes of other measurements within its future light cone. 
This gives it very peculiar properties, which evoke the 
suspicion that it must already be excluded by experiments
other than Bell experiments and/or cosmological observations. 
Perhaps it is, but this has not yet been shown. 
It is an interesting challenge to our understanding of physics 
to seek conclusive evidence for quantum theory against
this peculiar and even more counter-intuitive alternative.  

I first review the definition of causal quantum theory,
its theoretical motivation, and its very radical
implications.  
I then review the evidence from Bell experiments. 
As is well known, a long sequence of successively
more sophisticated Bell experiments have confirmed the predictions of standard quantum
theory.  However (with one only partial exception discussed below)
all Bell experiments to date are also consistent 
with causal quantum theory if the collapse process 
only takes place when macroscopic displacements of masses
or significantly different gravitational fields are
superposed.  They are also consistent with causal
quantum theory if collapse requires a measurement outcome enters
the consciousness of an observer.  
Since these are the most popular collapse hypotheses,
causal quantum theory is not definitively excluded
by Bell experiments to date.  

Finally, I consider other experiments that could
exclude causal quantum theory. 

\section{Causal quantum theory} 

Quantum theory and special relativity are related rather subtly and 
beautifully, in a way that allows quantum theory to violate
local causality without allowing superluminal signalling.
There is nothing evidently problematic in this relationship,
except perhaps that it is not so obvious how to extend it
to include a theory of gravity.  
Still, it is interesting to consider alternatives, even if ultimately only
to be sure we understand just how compelling the evidence is for standard
quantum theory.  

Causal quantum theory assumes that measurements or collapses 
are well-defined physical processes, localised in space-time,
and never give perfectly reliable or definitive outcomes.
It then ensures consistency with special relativity
by postulating that measurement or collapse outcomes respect a strong
form of Minkowski causality, in that a measurement's outcome only
influences the probability distribution for the outcomes of
other measurements within its future light cone. 
There may be no strong reason to prefer these assumptions
over those of standard quantum theory, even in the absence 
of experimental evidence.
However, they are all reasonably well motivated: 
the following paragraphs sketch some reasons.  

Collapse hypotheses can be motivated as solutions to the 
quantum reality (or measurement) problem, as alternative routes to
unifying quantum theory and gravity without necessarily quantising
gravity in any standard sense, or even as speculative ways of
connnecting consciousness and physics.  All of these motivations
are questionable, but all have thoughtful proponents.      

If collapses are objective, it is quite plausible that they
are typically well localized events, and indeed this is 
a feature of some explicit collapse models \cite{ghirardi1986unified,ghirardi1990markov}. 
As far as we know, all real world measurements are imperfect,
and it seems very plausible that this is true of all possible
real world measurements or observations, including those made
by our organs of perception and brains. 
It is also a feature of most well known explicit collapse 
models that collapse outcomes are never definitive -- ``tails''
in the wave function persist.  
Also, as far as we know, quantum states in the real world
never precisely lie in the kernel of real world measurement
operators.  

The strong form of Minkowski causality that defines
causal quantum theory holds in classical physical models
respecting the relativity principle, and the similarly
strong version of Einstein 
causality holds in general relativity.     All else being
equal, imposing this version of causality on a physical theory
is the simplest way to ensure that it respects at least
the signalling constraints implied by special and
general relativity.

\subsection{Measurement model and assumptions} 

We can define {\it causal quantum
theory} \cite{kent2005causal} by an abstract
black box model of measurements, rather than
specifying a particular localized collapse theory
or measurement model.  
We will focus here on measurements that are approximations
to an ideal detection of a single particle: it is easy
to generalize our discussion to other types of 
local measurements.   
In our model, a measurement $M$ is defined by a set of 
Kraus operators $\{ A_i \}_{i \in I}$ and is 
a physical operation that takes place inside a 
finite black box, which produces an output (the measurement outcome $i$).
To simplify, we suppose the box is of negligible size 
and that the measurement takes negligible time,
so that in our model we can approximate 
the measurement output as being created at a definite point $x_M$ in space-time.
We assume a fixed background space-time with no closed time-like
curves.  Our discussion applies in any such space-time, but
for definiteness we consider Minkowski space unless otherwise specified.  

The measurement operators include a distinguished operator, $A_0$, 
whose outcome is supposed to correspond approximately to the event that no particle
was detected in the box.   The other outcomes $i \neq 0$
are supposed to correspond approximately to a detection of a particle
in the box, perhaps together with other information -- for example,
the type of particle detected and/or information about internal degrees
of freedom.   So, suppressing internal degrees of freedom to 
simplify the illustration, if $\psi (x_1 , \ldots , x_n )$ is an $n$-particle
state with zero particle probability density inside the box, 
we have $A_i \psi \approx 0$ for $i \neq 0$.\footnote{
For simplicity, we assume here the particles are distinguishable, but that the 
detector in this illustration does not distinguish them.
In general, we allow any type of detector that locally detects
particles, including detectors that detect some types but not others.   
It is also straightforward to 
deal with bosons or fermions, at the price of slightly complicating
the notation.
In principle, our definitions are also meant to extend to relativistic quantum
fields, modulo the standard problems in rigorizing their 
measurement theory.} 

The measurement operators $A_i$ are supposed to roughly approximate
projections, $A_i \approx P_i$, where $P_i$ is a projection 
operator corresponding to the relevant outcome.  
The sense in which this approximation holds depends on the
details of the collapse or measurement model in question.    
A good illustration is a GRW \cite{ghirardi1986unified} localization operator $A^a_{x_0}$, which acts on single
particle wave functions $\psi (x)$ by 
\begin{equation}
\psi (x) \rightarrow C \exp ( (x- x_0 )^2 / a^2 ) \psi (x) \, ,
\end{equation}
where $a$ is a constant of the localisation model and $C$
is a normalisation constant.   
This approximates a projective measurement of position onto 
the interval $ \left[ x_0 - a , x_0 + a \right]$.   

While the Kraus operators may correspond to
the specified outcomes to very good approximation, the simplest
version
of our model requires
that none of them has a zero
eigenstate.  For example, for any $\psi$ as above and any $i$, 
$A_i \psi \neq 0$, although $ | A_i \psi |$ may be very small.  
GRWP \cite{ghirardi1986unified,ghirardi1990markov}
spontaneous localisation model collapses have this feature. 

One way to interpret this in models of measurement is to  think of the measurements as carried out
by imperfect detectors that always have some nonzero probability
of giving false positive or negative detections, of 
misidentifying the particle type, and so on.   As far as we know,
every real world detector, including the perception organs
and brains of conscious observers,
indeed has these properties. 

A more general version of our model requires only the weaker
condition that, given any physically realisable initial state $\psi_S$
defined on a hypersurface $S$ prior to space-time points
$x_1 , \ldots , x_n$, 
the prescribed unitary evolution law, and any sequence of measurement
operators $A^1_{i_1} , \ldots A^n_{i_n}$ localised near the space-time
points $x_1 , \ldots x_n$, the final state $\psi'_{S'}$  arising
on any hypersurface $S'$ to the future of $x_1 , \ldots , x_n$ 
is well-defined (i.e. non-zero). 
This could be justified by noting that in practice we can 
never produce pure states of a precisely specified form, 
and every physically realised state contains small 
uncontrolled components that are not annihilated by any
possible sequence of measurements.   This may also be true of the
cosmological initial state. 

The effect of obtaining outcome $i$ is, as usual, to replace
an initial state $\rho$ by the
post-measurement state 
\begin{equation}\label{collapse}
 A_i \rho A_i^{\dagger}/ \Tr ( A_i \rho
A_i^{\dagger} ) \, .  
\end{equation}
We allow here the possibility that $\rho$ may be mixed.    
Since the Kraus operators have no zero 
eigenstates, the denominator is always nonzero, and so 
this effect is always well defined.  

So far, this picture is
consistent with standard formulations of quantum theory, 
modulo some vagueness about whether we can
think of measurement outcomes as arising at a definite space-time
point (and if so exactly which point).   
In a version of Copenhagen quantum theory with prescribed Heisenberg
cuts defining measurement apparatus, we 
can think of our black boxes as defined spatially by measurement apparatus
and temporally by the measurement process, or by regions within these
space-time regions.   On Wigner's hypothesis, we can define the black
boxes spatially by the brains of conscious observers and temporally
by their conscious perception times.   

The Dirac-von Neumann treatment of measurements corresponding
to projection operators can be understood in this model -- as
generally in modern formulations of quantum theory -- as a mathematically
useful but physically unrealistic limiting case.  

\subsection{Standard relativistic quantum theory}

Now suppose that we have a relativistically covariant quantum
evolution law, with quantum states $\psi_S$ defined on space-like
hypersurfaces $S$ via the Tomonaga-Schwinger formalism.   
We need to include the effects of measurements within this
framework.   

The standard way to do this
is to define $\psi_S$ by applying the unitary evolution
from the initial hypersurface $S_0$ to $S$, together with 
the measurement transformations (\ref{collapse}) for all
measurements taking place between $S_0$ and $S$.   
The outcome probabilities of measurements on (or in the near 
future of) $S$ are then obtained from $\psi_S$ in the usual
way.  In particular, the outcome probabilities for a measurement
localized at the point $x \in S$ depend only on the local 
density matrix of $\psi_S$ at $x$. 
Since the quantum measurement postulate is consistent with
Minkowski causality, this prescription gives a well-defined
answer and is Lorentz covariant.   In particular, it 
does not matter in which order (\ref{collapse})
is applied for space-like separated measurement events.
This prescription defines standard relativistic quantum theory within
our model of measurements.  

\subsection{Causal quantum theory}

By contrast, in causal quantum theory, the outcome probabilities for a measurement
localized at the point $x$ depend only on the local 
density matrix of $\psi_{\Lambda(x)}$ at $x$, where, loosely
speaking, $\psi_{\Lambda(x)}$ is the wave function defined on
the past light cone $\Lambda (x)$ of $x$.   More precisely, 
the local density matrix is the limit of the local
density matrices for the wave functions of spacelike
hypersurfaces $S$ tending to $\Lambda (x)$.   
Equivalently, we calculate the local density matrix
at $x$ from $\psi_S$, defined for any spacelike hypersurface
$S$ through $x$, but for this calculation we define $\psi_S$
allowing only for the outcomes of measurements inside
$\Lambda( x )$.   According to this prescription, if $i$ and $j$ 
are possible outcomes of measurements at spacelike separated
points $x$ and $y$, then $\Prob (i) = \Prob ( i | j )$. 
In other words, conditioning on space-like separated measurement
events makes no difference.

This prescription, clearly, is {\it not} 
consistent with standard quantum theory.   
For example, it predicts that if we can arrange
spacelike separated measurement events
that correspond approximately to spin measurements
of two separated particles in a spin singlet, 
the outcomes will be random (as in standard quantum 
theory) but also uncorrelated, whatever measurements
are chosen.   
Standard quantum theory predicts approximate anti-correlation
when the measurements are the same.  We now consider this in
more detail. 

\subsubsection{Internal consistency of causal quantum theory}

Is this prescription
self consistent?  At first sight it may seem that spacelike
separated measurements at $x$ and $y$ can give inconsistent outcomes. 
For example, spacelike separated $s_z$ measurements on a 
singlet state 
\begin{equation}\label{singlet}
\Psi_{-} = 
 ( { {1} \over { \sqrt{2}}} )
  ( \ket{\uparrow}_L \ket{\downarrow}_R  - \ket{\downarrow}_L \ket{\uparrow}_R )
\end{equation}
could give outcomes
coresponding to $\ket{\uparrow}_L$ and $\ket{\uparrow}_R$. .      
If this were precisely true, the quantum state and local density matrix would be undefined
at points in the joint future of $x$ and $y$, and we would have no
prescription for the probabilities of measurement outcomes there.  

Recall, though, that our model requires that the Kraus operators defining
measurement outcomes have no zero eigenvalues.   However closely a product of such operators approximates a
projection operator, it cannot annihilate the quantum state.
The resulting state may have very small norm, but 
the denominator in (\ref{collapse}) ensures a normalised
post-measurement state 
after any sequence of measurement outcomes.   

We could also invoke the fact that, as far as we know, it is
impossible to prepare or find in nature a system represented
precisely by the state (\ref{singlet}).  
A full discussion would take this into account, 
noting that at best we can prepare a mixed state
dominated by states of the form 
\begin{equation}\label{mixed}
\Psi =    a_{\uparrow \downarrow} \ket{\uparrow}_L \ket{\downarrow}_R
+ a_{\downarrow \uparrow} \ket{\downarrow}_L \ket{\uparrow}_R +
a_{\uparrow \uparrow} \ket{\uparrow}_L \ket{\uparrow}_R
+ a_{\downarrow \downarrow} \ket{\downarrow}_L \ket{\downarrow}_R  \,
, 
\end{equation}
where the four coefficents are all non-zero, with
$ a_{\uparrow \downarrow}  \approx  { {1} \over { \sqrt{2}}} , 
a_{ \downarrow \uparrow}  \approx  - { {1} \over { \sqrt{2}}} , 
a_{\uparrow \uparrow} \approx 0 , a_{\downarrow \downarrow} \approx
0$.  For simplicity of illustration, we neglect this 
here (although it is an important defence of causal
quantum theory and the details may significantly affect
its implications in any given example) and focus on the first point. 

Suppose then that the state of the relevant system is 
precisely (\ref{singlet}).   The operator corresponding to 
the outcome we label ``$s_z = \ket{\uparrow}$'' might
for example take the form 
\begin{equation} 
A^{\uparrow} = (1 - \epsilon)^{1/2} \ket{\uparrow} \bra{\uparrow}
+ \epsilon^{1/2} \ket{\downarrow} \bra{\downarrow} \, ,
\end{equation}
and similarly the outcome we label ``$s_z = \ket{\downarrow}$'' is
\begin{equation}
A^{\downarrow} = (\epsilon)^{1/2} \ket{\uparrow} \bra{\uparrow}
+ (1 - \epsilon)^{1/2} \ket{\downarrow} \bra{\downarrow} \, .
\end{equation} 
Our model allows $\epsilon$ to be arbitrarily small, but requires
$\epsilon > 0$; we take $\epsilon \ll 1$. 

We then have that
\begin{equation}
A^{\uparrow}_L A^{\uparrow}_R \Psi_- = 
{{1} \over { \sqrt{2}}} (1 - \epsilon)^{1/2} \epsilon^{1/2}  ( \ket{\uparrow}_L
\ket{\downarrow}_R
 - \ket{\downarrow}_L \ket{\uparrow}_R \, ,
\end{equation}
which has norm $\epsilon ( 1 - \epsilon )$ and normalises to $\Psi_-$.   
The key point here is that, while the norm is small,
it is non-zero.   Hence, we still have a well-defined state and 
a consistent prescription for obtaining measurement probabilities
in the joint future of $x$ and $y$.   

Now consider an extension of this experiment, in which the 
L and R wings are widely separated enough that a sequence 
of $N$ measurements can be carried out within a region $R_L$
on the L wing, and a sequence of $N$ measurements within
a region $R_R$ on the R wing, with the regions 
$R_L$ and $R_R$ space-like separated. 

Suppose that the first measurement on the L wing  produces
outcome $A^{\uparrow}_L$.   According to causal quantum
outcome $A^{\uparrow}_L$.   According to causal quantum
theory, the relevant unnormalised state for calculating outcomes
of the second measurement on the L wing is
\begin{equation}
A^{\uparrow}_L \Psi_- = 
{{1} \over { \sqrt{2}}} (1 - \epsilon)^{1/2} ( \ket{\uparrow}_L
\ket{\downarrow}_R
 - \epsilon^{1/2} \ket{\downarrow}_L \ket{\uparrow}_R \, . 
\end{equation}
This normalises to 
\begin{equation}
(1 - \epsilon)^{1/2} ( \ket{\uparrow}_L
\ket{\downarrow}_R
 - \epsilon^{1/2} \ket{\downarrow}_L \ket{\uparrow}_R \, . 
\end{equation}
The second measurement thus has outcome $A^{\uparrow}_L$ with
probability 
\begin{equation}
( 1 - \epsilon )^2 + \epsilon^2 = 1 - 2 \epsilon + 2 \epsilon^2
\end{equation}
Assuming this outcome is realised, the third measurement has
the same outcome with probability
\begin{equation}
  1 - \epsilon - \epsilon^2 + O( \epsilon^3 ) \, ,
\end{equation}.
If all previous measurements produced outcome $A^{\uparrow}_L$,
each successive measurement produces the same outcome with
conditional probability closer to $(1 - \epsilon )$. 

Given that the first outcome is $A^{\uparrow}_L$,
with high probability, the first several outcomes will be 
$A^{\uparrow}_L$, making the local state closer and closer to 
$ \ket{\uparrow}_L \bra{\uparrow}_L$.   Outcomes $A^{\downarrow}_L$ will continue to have probability
$\epsilon + O( \epsilon^2 )$, and will be realised with frequency
$\approx \epsilon$.   Nonetheless, with probability close to $1$,
the local state will tend asymptotically to $ \ket{\uparrow}_L
\bra{\uparrow}_L$.   A sequence of $N$ measurements with
proportion  $\approx (1- \epsilon)$ having outcome $A^{\uparrow}_L$
will give local observers on this wing increasingly strong
evidence that, for practical purposes, they may take their
local state to be very close to $ \ket{\uparrow}_L
\bra{\uparrow}_L$, {\it so long as measurement outcomes elsewhere
may be neglected}.

Now suppose the first measurement on the R wing also  produces
outcome $A^{\uparrow}_R$.   This will be the case with probability
$\frac{1}{2}$, independent of the outcomes of the measurements
on the L wing.  The same discussion applies.  So, with
probability close to $1$,
the local state on the R wing will tend asymptotically to $ \ket{\uparrow}_R
\bra{\uparrow}_R$, and local observers will obtain increasing
evidence that, for practical purposes, they may take their
local state to be very close to $ \ket{\uparrow}_R
\bra{\uparrow}_R$, again so long as measurement outcomes elsewhere
may be neglected.

\begin{figure}[h]
\centering
\includegraphics[width=\linewidth, height=10cm]{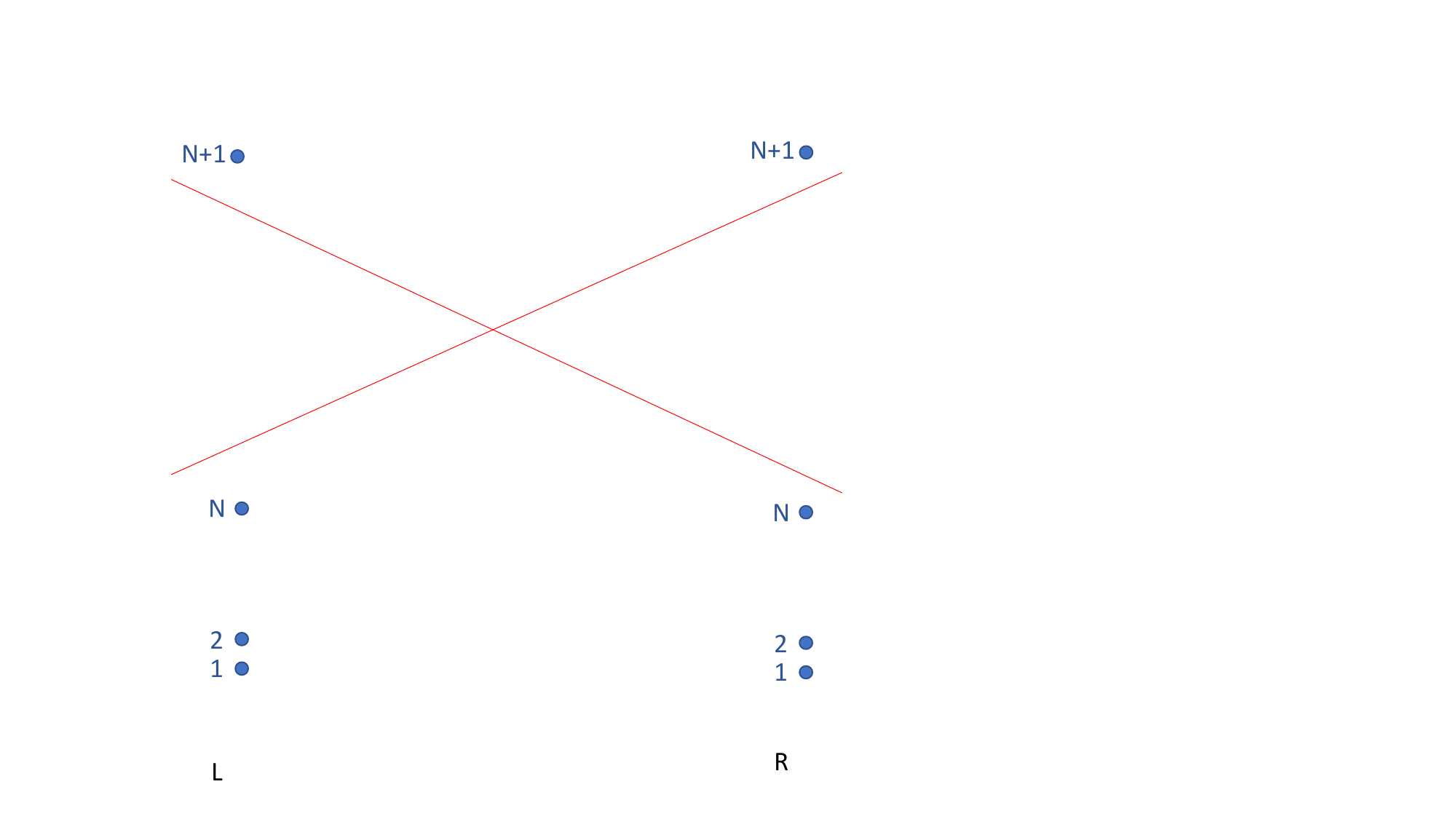} 
\caption{Spacelike separated approximate measurements on a singlet
in causal quantum theory.    The observers in each wing carry out
$N$ measurements with the two sets of measurements in space-like
separated regions.   They then both wait until the measurements 
from the other wing are in their past light cone  before carrying
out a further measurement.   Measurements are denoted by blue
dots and light rays by red lines.}
\label{nmments}
\end{figure}

Now suppose both observers pause after $N$ measurements, and 
both wait until all the measurements from the other wing 
are in their past light cone.   At this point their local
states revert to $\frac{1}{2}I_L$ and $\frac{1}{2}I_R$  
respectively.   Whether or not they are aware of the 
measurement outcomes in the other wing, their
outcome probabilities for their next measurement change:
according to causal quantum theory, $A^{\uparrow}_L$ and
$A^{\uparrow}_R$ now
both have probability
$\frac{1}{2}$ (see Fig. \ref{nmments}). 

This is a general feature of causal quantum theory: evidence
about the state of a local system determines the future
behaviour of that state only insofar as measurement outcomes
elsewhere remain causally separated.   This is true no matter  
how compelling that evidence would be in standard quantum theory.
And it is true for systems of any size or complexity.  
In principle, a dynamical collapse model
\cite{ghirardi1986unified,ghirardi1990markov}
version of causal quantum
theory could allow an apparently stable galaxy obeying quasiclassical
laws to persist for a long period, only to be ``overwritten'' in 
the future when its past light cone includes
currently space-like separated measurement outcomes that are
inconsistent
with its state.\footnote{
As Marx might have put it, all that is solid melts into air.}
Inhabitants of such a galaxy who know they live in a universe
described by causal quantum theory might (depending on the 
details of the relevant collapse or measurement hypothesis)
be motivated to frantically carry out more and more measurements
confirming its and their existence, and to broadcast the data in
the hope that it might be measured again elsewhere, with the aspiration of out-weighting
any currently space-like separated and inconsistent measurement 
data.\footnote{There are analogies in present-day politics, alas.}

\subsection{Is causal quantum theory too strange to contemplate?}

Causal quantum theory might be criticized as maintaining internal
consistency only by invoking the possibility of highly improbable
measurement errors and/or improbable outcomes arising from wave
function tails.     
This is circular reasoning, though. 
{\it According to standard quantum theory},
the measurement errors and outcomes that routinely arise in causal quantum
theory are indeed highly improbable. 
However, causal quantum theory is a new theory
with different rules for the probability of a sequence of
measurement outcomes, and according to causal quantum theory
its measurement outcome predictions are not improbable.  
An unbiased scientific
comparison of standard quantum theory and causal quantum theory
cannot invoke statements that apply
only within one theory as reasons for disbelieving
the other.   

Again, it is worth emphasizing that, however strange causal quantum theory may be,
it does have a respectable theoretical motivation.  To recap, this runs as follows.
First, we accept the empirical success of non-relativistic 
quantum theory when dealing with individual systems, and take
this as evidence that the basic mathematical formalism of quantum
theory is an essential part of the description of nature.  
Second, we note that measurement plays a key role in standard
quantum theory.
We note too that an adequately general and realistic account
of quantum measurements is given by considering Kraus operators 
with no zero eigenvalues, and that it is not absurdly 
unreasonable to postulate
that all measurements are defined by such operators.  
Third, noting the EPR argument, and the a priori puzzling relationship
between quantum theory and special and general relativity, we make
the assumption that measurement outcomes influence one another
in a way that respects strict Minkowski and Einstein causality.  That is, 
any given measurement outcome only affects events in its causal
future. 
Although this last assumption is non-standard, it seems {\it a priori} 
quite natural within the framework of special and general relativity.  
Causal quantum theory is the result.

One could even imagine a counter-factual 
history in which the theory of approximate measurement was 
developed soon after 1926, 
and the idea that measurement might be a definitely
localized physical process was taken seriously from
the start.   Given the then unresolved tension between quantum theory
and special relativity, causal quantum theory might 
conceivably have been proposed as a 
logically consistent possibility in, say, 1930.  
After the EPR argument was presented, one could
just about imagine causal quantum theory and standard
quantum theory initially being seen as genuine competitors, which needed
to be distinguished empirically.  

Of course, history followed another path.   The EPR argument
ultimately led to Bell's theorem and Bell experiments, which
are generally taken to be compelling confirmations 
of standard quantum theory and Bell non-locality.  
The theory of approximate
measurements was developed much later than the EPR argument.  
And, although the idea that measurement is a localized 
physical process is quite well aligned with ideas explored by
some of the founders of quantum theory, it was not
generally seen as an idea that might have testable implications
worth exploring until the work of Ghiradi-Rimini-Weber-Pearle
\cite{ghirardi1986unified,ghirardi1990markov}, Diosi \cite{diosi1987universal},
Penrose \cite{penrose1996gravity} and others.  

\subsection{What is the fundamental status of causal quantum theory?}

The extent to which ordinary quantum theory is
well-defined is already a delicate question.   Some standard
textbook versions of quantum theory offer versions
of a Copenhagen interpretation, in which measuring
devices or macroscopic amplifications play a fundamental
role in defining the theory.    No precise definition
of measuring device or macroscopic is available, and
so there is an ineliminable vagueness in principle in
these versions of quantum theory.   Wigner's speculative
idea \cite{wigner1961remarks} consciousness
plays a fundamental role, and is essential to the definition 
of measurement, can be thought of as another Copenhagen-like
interpretation.   Again, though, there is an ineliminable
vagueness, since we lack any precise model of consciousness.

If we take these interpretations as requiring some 
qualitative distinction between macroscopic and microscopic,
or conscious and unconscious, then they imply that quantum
theory is not universal.   Alternatively, if we try to
set out an Everettian version \cite{saunders2010many} in which quantum theory
is universal, we run into problems of structure, 
probability and the problem of theory confirmation
for many-worlds theories \cite{kent2010one}, and we 
find a lack of consensus amongst many-worlders on the
answers.    

Similarly, we can consider Copenhagen-like versions of causal quantum
theory, in which measuring devices (or conscious minds) are 
taken to be qualitatively distinct from quantum systems.  
In these versions, the output of a measuring device is 
treated as a stable classical record.   However, because
we assume that measurements are necexssarily imperfect, 
successive measurements on a quantum system do not 
necessarily produce the same output.

Alternatively, we can think of causal quantum theory as 
a causal version of some form of objective collapse theory
\cite{ghirardi1986unified,ghirardi1990markov}, with 
localised collapses, that
aspires to give a unified treatment of microscopic
and macroscopic physics.     
Dynamical collapse models are designed to resolve 
the problems of Everettian quantum theory by 
postulating objective probabilities that effectively
single out one quasi-classical world as realised from among infinitely
many possible such worlds.   
Whether they completely succeed is debated \cite{sep-qm-collapse}.
Fully satisfactory relativistic collapse models have proven elusive.
Also, of course, collapse models make testably distinct predictions
from standard quantum theory, and may be refuted
\cite{bassi2013models}. 

Causal quantum theory is thus an umbrella term for a class
of theories.   While this is true of standard quantum theory
in a sense, it is true of causal quantum theory in a
stronger sense: causal quantum theory needs some localized collapse
hypothesis, and different localized collapse hypotheses can make very
different predictions in causal quantum theory, even when they are
effectively indistinguishable when applied to standard quantum theory.   

\section{Empirical tests of causal quantum theory} 

\subsection{Stronger Bell experiments}

Causal quantum theory can be thought of as a local hidden
variable theory, in which the local hidden variable is
the local quantum state, and is influenced by measurement
outcomes in its past light cone.    
Any given version of causal quantum theory can thus be excluded 
by Bell experiments that close the relevant version of the
collapse locality loophole, by ensuring that, according 
to the relevant collapse or measurement model, space-like
separated measurement actually take place in the two wings. 

Indeed, causal quantum theory is somewhat easier to exclude
than a generic local hidden variable theory exploiting
the same version of the collapse locality loophole, since
it makes specific predictions that differ from those
of standard quantum theory even for experiments with
a single measurement choice on each wing. 
For example, as discussed above, causal quantum theory
predicts that, if we prepare an approximate singlet state 
\begin{equation}
\ket{\psi} \approx 
\frac{1}{\sqrt{2}} ( \ket{\uparrow}_A \ket{\downarrow}_B -
\ket{\downarrow}_A \ket{\uparrow}_B ) \, , 
\end{equation}
and ensure space-like separated measurements of $s_Z$ on 
each wing, the outcomes in the two wings will be 
uniformly random and uncorrelated:
\begin{equation}
P_{\rm causal~qt}  ( \uparrow_A , \uparrow_B ) \approx
P_{\rm causal~qt}  ( \uparrow_A , \downarrow_B ) \approx
P_{\rm causal~qt}  ( \downarrow_A , \uparrow_B ) \approx
P_{\rm causal~qt}  ( \downarrow_A , \downarrow_B ) \approx 1/4 \, , 
\end{equation} 
whereas 
\begin{equation}
P_{\rm standard~qt}  ( \uparrow_A , \downarrow_B ) \approx 
P_{\rm standard~qt}  ( \downarrow_A , \uparrow_B ) \approx 1/2 \, , \qquad  
P_{\rm standard~qt}  ( \uparrow_A , \uparrow_B ) \approx 
P_{\rm standard~qt}  ( \downarrow_A , \downarrow_B ) \approx 0 \, . 
\end{equation} 

There has been some progress in this area since causal quantum
theory was first described \cite{kent2005causal}.   In particular,
motivated by these ideas, 
Salart et al. \cite{salart2008spacelike} carried out a Bell experiment closing the
collapse locality loophole assuming that collapses 
arise to prevent superpositions of distinguishable 
gravitational fields and can be characterised by 
quantitative guesstimates due to Diosi and Penrose. 
This experiment refuted a version of causal quantum
theory based on the same assumptions.
It left open the question of whether causal 
quantum theory could still hold if Diosi and Penrose's
estimates of collapse time were increased by a 
factor of $\approx 10^2$, or other versions of 
gravitational collapse model were implied, or 
other collapse or measurement models were assumed. 
Techniques for implementing stronger Bell
experiments that should be able to test
the collapse locality loophole for most
interesting collapse models were described in
Ref. \cite{kent2020stronger}.   Such experiments
could also refute causal quantum theory.   

\subsection{Testing for multiple detections of 
a single particle}

Recall that our black box model of measurement assumes 
some characterisation of measurement or collapse-inducing devices,
which may for example involve reconfiguring mass distributions or 
perceptions in human brains.  
Suppose that these
devices detect whether or not a particle is in the black box,
and have efficiency $(1 - \epsilon)$, in the following sense.
The measurement for
box $i$ is defined by two operators $A_0^i , A_1^i$. 
Let $\psi_0^i$ be any single particle 
state with support outside the box, and $\psi_1^i$ 
be any single particle state with support inside the box. 
Let $\rho_0^i$ and $\rho_1^i$ be the corresponding density matrices.  
Then we assume 
\begin{eqnarray}
\Tr ( A_0^i \rho_0^i ( A_0^i )^{\dagger} ) &=& 1 - \epsilon \,  , \\ 
\Tr ( A_1^i \rho_1^i ( A_1^i )^{\dagger} ) &=& 1 - \epsilon \,  , \\ 
\Tr ( A_0^i \rho_1^i ( A_0^i )^{\dagger} ) &=& \epsilon \,  , \\ 
\Tr ( A_1^i \rho_0^i ( A_1^i )^{\dagger} ) &=& \epsilon \, . 
\end{eqnarray}
We also assume that
$A_0^i \psi_1^i$ and $A_1^i \psi_1^i$ have support inside the box,
and that 
$A_0^i \psi_0^i$ and $A_1^i \psi_0^i$ have support inside the 
same region as $\psi_0^i$ (and so, in particular, have support 
outside the box).  

Thus, in a single experiment with a single measurement
box $i$ on a single particle state with support inside the box,
the detector will click with probability $1 - \epsilon$; if the state
has support outside the box, the detector will click with
probability $\epsilon$.  

Now consider an experiment with $N$ spacelike separated
(hence disjoint) boxes, labelled by $1 \leq i \leq N$, on a quantum state
$$
\psi = {{1} \over {\sqrt{N}}} ( \sum_i \psi_1^i ) \, ,
$$
where the $\psi_1^i$ are normalised states with support in box $i$, as
above.  According to standard quantum theory, if we obtain a click
from box $i$, the post-detection state is 
\begin{eqnarray}
\psi'_i &=& { A_1^i \psi  } \over { | A_1^i \psi | }  \\
& = & (1 + (N-2) \epsilon)^{-1/2} ( A_1^i \psi_1^i  + \sum_{j \neq i} A_1^i \psi_1^j ) \, .
\end{eqnarray}
Here the normalisation factor is obtained
using that the states $A_1^i \psi_1^j $ for $1 \leq j \leq N$ are all
orthogonal, with $ | A_1^i \psi_1^i |^2 = 1 - \epsilon$
and $ | A_1^i \psi_1^j |^2 = \epsilon$.   
The probability $P_{QM} (i)$ of getting a click in any given box $i$ is thus
roughly $N^{-1}$; the conditional probability $P_{QM} (j|i)$
of getting a click in box $j$ given that we get one in a specified box
$i$
is $O( \epsilon )$.
More generally, the conditional probability
$P_{QM} (k|i_1 \ldots i_m)$, where $k$ is distinct from $i_1 , \ldots
, i_m$, is also $O( \epsilon )$.  
To $O( \epsilon)$ the probabilities $P_{QM} (n) $ of getting $n$
clicks are given by
$P_{QM} ( 1 ) = 1$, $P_{QM}(n ) = 0 $ for $n \neq 1$.

In causal quantum theory, on the other hand, so long as the
measurement events are spacelike separated, the probability
of each outcome is independent of the others.
We have $P_{CQM} (i) = P_{CQM} ( j | i ) = P_{CQM} (k | ji ) = \ldots
\approx N^{-1}$.   The probabilities $P_{CQM} (n) $ of getting $n$
clicks obey
\begin{equation} 
P_{CQM} ( n ) \approx \frac{1}{n!} e^{-1} \, ,
\end{equation}
for $n \ll N$.  

The expectation value of the number of clicks is $1$ in both
theories. 
However, the variances are different.  Causal quantum theory predicts multiple and zero 
click outcomes that are very unlikely according to standard quantum theory.
Note that causal quantum theory does {\it not} 
predict that, when multiple detections are made, the system goes
on indefinitely to behave as though it now contains multiple
particles. 
Events in the joint future of all
the detectors are predicted by allowing for all the detector
measurements, in a way consistent with the
possibility that all but one of the detectors produced a false
positive result -- a possibility which always has nonzero
probability.   Trying to bring the ``detected particles'' 
together in order to verify violation of a conservation law
(which has zero probability in standard quantum theory)
will thus always fail. 

Nonetheless, spacelike separated sets of detectors should often register anomalous multiple detections from
a single particle, and in our measurement model evidence of such multiple
detections persists even in the intersection of their future light
cones.   It is interesting to ask whether we should expect
to have seen evidence of something like this in existing experiments, observations such
as cosmic ray tracks in mica, or elsewhere, if causal quantum theory were correct.
It is not obvious that we should, given that the expectation value of the number of 
detections is the same in causal quantum theory as in standard quantum
theory.   One reason is that our measurement model
does not always apply: the probability of a track appearing in mica in
the absence of a cosmic ray is likely astronomically small. 
Another more generally relevant reason is that, as in 
Bell experiments, the prediction requires specifically 
designed space-like
separated measurement devices appropriate for the relevant
collapse postulate. 

The case of $N=2$ suffices for experimental tests distinguishing
causal and standard quantum theory, since the rate of anomalous
double clicks would be high ( $\frac{1}{4} + O (\epsilon )$) according
to causal quantum theory and low ($O( \epsilon )$) according to 
standard quantum theory.   Causal quantum theory can thus
be tested with a single photon source and beamsplitter
together with appropriate space-like detectors.
This makes it somewhat easier to refute causal quantum theory than
to close the collapse locality loophole in general: 
the latter requires a source of entangled particles 
as a component of the experiment, although not
necessarily long range entanglement distribution \cite{kent2020stronger}.

A very nice experiment of this type 
was carried out and carefully analysed by Guerreiro et al.\cite{guerreiro2012single}
In their experiment, single heralded photons were observed at one (and
always only one) of two photodetectors, arranged so that the
possible detection events were necessarily spacelike separated.  
Unlike the experiment of Salart et al. \cite{salart2008spacelike}, 
discussed above, these detection events were not macroscopically
amplified in spacelike regions.   The Guerreiro et al. experiment
thus did not test a version of causal quantum theory associated
with any particularly plausible of well known localized collapse
model.    However, the techniques of Refs.
\cite{kent2020stronger,guerreiro2012single,
salart2008spacelike} could certainly be combined, and extended
to other collapse hypotheses.   Indeed, we hope this discussion
will stimulate further experimental work in this area. 

Causal quantum theory thus can and very likely will be refuted by
experiment.   Nonetheless it is a reasonably well motivated radical
alternative to standard relativistic quantum theory that has
essentially gone unnoticed for many decades and quite plausibly
still awaits conclusive refutation.  
It raises the question: what else might we be missing?

\section{Conclusions}

Causal quantum theory relies on 
some form of objective localized collapse or measurement
model.  These models may of course be incorrect, and their
motivations, though reasonable, are not universally appreciated.
Causal quantum theory is also a strange theory that makes
predictions that run counter to the intuition of anyone
familiar with standard quantum theory.   
It may perhaps already be contradicted by experiment or
observation, though it does not seem obvious that it is. 

Different people may reasonably assign different weights
to each of these arguments, but all of them give some reason
to be sceptical about causal quantum theory, and in 
combination they probably give very strong reasons to
be sceptical.   Even so, a conclusive experimental 
refutation would be more compelling, and I hope 
the present discussion may stimulate experimental
work.  

One might, perhaps, also entertain the 
idea that causal quantum theory could apply to 
some part of nature but not all.   The most
obvious candidate here is the gravitational field,
given that Einstein causality is a fundamental
feature of general relativity.   A theory in 
which the gravitational field is determined
by causal quantum theory rules, while matter
correlations follow standard quantum predictions, 
might seem even stranger than fully-fledged
causal quantum theory.  Again, though, a 
conclusive experimental refutation would be 
a more compelling argument.   

\vskip10pt

{\bf Competing Interests} \qquad There are no competing interests. 

\vskip10pt

{\bf Data accessibility} \qquad This work does not have any
experimental data.

\vskip10pt
{\bf Ethics statement} \qquad This work did not involve any active
collection of human data
\vskip10pt
{\bf Funding statement} \qquad This work was partially 
supported by a grant from FQXi and by
Perimeter Institute for Theoretical Physics. Research at Perimeter
Institute is supported by the Government of Canada through Industry
Canada and by the Province of Ontario through the Ministry of
Research and Innovation.  
\vskip10pt
{\bf Acknowledgements} \qquad I thank Jess Riedel, Nicolas Gisin and
Renato Renner for helpful
comments and Gerard Milburn for pointing out an error in an
earlier draft.

\bibliographystyle{unsrtnat}
\bibliography{collapselocexpt}{}
\end{document}